\documentclass[]{interact}

\usepackage{epstopdf}
\usepackage[caption=false]{subfig}
\usepackage[numbers,sort&compress]{natbib}
\bibpunct[, ]{[}{]}{,}{n}{,}{,}

\theoremstyle{plain}

\theoremstyle{definition}

\theoremstyle{remark}

\begin{document}


\title{Effects of a {\bf k}-dependent Hybridization on the Fermi Surface of an Extended $d-p$ Hubbard Model}

\author{
\name{D. M. Lalis\textsuperscript{a}, E. J. Calegari\textsuperscript{a}\thanks{CONTACT E. J. Calegari. Email: eleonir@ufsm.br}, L. C. Prauchner\textsuperscript{a} and S. G. Magalhaes\textsuperscript{b}}
\affil{\textsuperscript{a}Departamento de F\'{\i}sica, Universidade Federal de Santa Maria, 97105-900, Santa Maria, RS, Brazil; \textsuperscript{b} Instituto de F\'{\i}sica, Universidade Federal do Rio Grande do Sul, 91501-970,
Porto Alegre, RS, Brazil}
}

\maketitle

\begin{abstract}
The topology of the Fermi surface of an extended $d-p$ Hubbard model is investigated using the Green's function technique in a n-pole approximation.
The effects of the $d-p$ hybridization on the Fermi surface
are the main focus in the present work. 
Nevertheless, 
the effects of doping, Coulomb interaction and hopping to second-nearest-neighbors on the Fermi surface, are also studied. 
Particularly, it is shown that the crossover from hole-like to 
electron-like Fermi surface (Lifshitz transition) is deeply affected by the $d-p$ hybridization. Moreover, the pseudogap present in the low doping regime is also affected by the hybridization. The results show that both the doping and the hybridization act in the sense of suppresses the pseudogap. Therefore, the systematic investigation of the Fermi surface topology, shows that not only the doping but also the hybridization can be considered as a control parameter for both the pseudogap and the Lifshitz transition. Assuming that the hybridization is sensitive to external pressure, the present results agree qualitatively with recent experimental data for the cuprate Nd-LSCO.

\end{abstract}


\section{Introduction}

More than three decades after the discovery of the high-temperature superconductors (high-$T_c$), 
the theoretical description of this phenomenon still represents a challenge for 
the physicists. 
To understand the nature of the pseudogap regime present in the normal state of the hole-doped cuprates, seems to be a essential ingredient for the process of understanding the superconductivity in these systems. There are experimental evidences suggesting a close relation between the existence of the pseudogap and the topology of the Fermi surface  in cuprates \cite{Doiron}. Both the pseudogap and the Fermi surface topology are strongly hole-doping dependent \cite{Doiron, Ino2}. When doping is increased, there is a critical doping, above which, the Fermi surface topology changes from hole-like to electron-like, indicating a Lifshitz transition \cite{benhabib,wei,helena}. The pseudogap occurs when the Fermi surface is hole-like. The origin of the pseudogap is still a subject of intensive debate in condensed matter physics. There are different theories proposed for the nature of the pseudogap \cite{proust}. Recent experimental \cite{chan,cyr} and theoretical \cite{proust,harrison,morinari18,morinari19} results suggest that the pseudogap emerges due to short-range antiferromagnetic correlations present in the underdoped strongly correlated regime of cuprates.

Due to the strong correlations at the Cu-sites,
the one-band Hubbard model has been largely used to describe such systems, leading to results which are important for elucidate the mechanisms of superconductivity in cuprates
 \cite{chen,helena,youhei,wei,fratino,foley}. Nevertheless, the fact that
the oxygen sites may be occupied by holes when the system is doped suggests that a model which also takes into account the oxygen can be more adequate to treat the doped regime of such systems \cite{calegari2005EB,xin,fratino2,kung,sidorov,vitali}.
In this sort of model, the hybridization between the $d$ orbitals of the cooper and the $p$ orbitals of the oxygen becomes an important parameter which is sensitive to changes of external pressure \cite{Sakakibara,aoki,Deng}. For the sake of simplicity,  it is usual to consider the hybridization as a $\vec{k}$-independent parameter \cite{calegari2005EB,sarasua,sengupta}. However, for a more appropriated description of these system, it is important to consider a $\vec{k}$-dependent hybridization \cite{macridin,mucio}.

 Motivated by recent experimental results relating the peseudogap and the Fermi surface topology of cuprate superconductors \cite{Doiron}, in the present work, a n-pole approximation \cite{Roth,beenen} has been used to investigate the topology of the Fermi surface associated with
an extended $d-p$ Hubbard model. The study of the Fermi surface and the dispersions of bands is essential
to better understand the superconducting and the normal properties of the high-$T_c$'s \cite{borisov}.
Our focus here,  is the normal state of a $d-p$ model where the pseudogap may occurs. It has been assumed a scenery in which short-range antiferromagnetic correlations are the source of pseudogap \cite{proust,chan,cyr,harrison,morinari18,morinari19}. 
The investigation of the effects of the doping and the hybridization on the topology of the Fermi surface is the central issue discussed in this paper.

\subsection{The model}

The Hamiltonian model is an improved version of the model studied in reference \cite{calegari2005EB}. It is given by:
\begin{eqnarray}
\hat{H}&=&\sum_{\langle\langle i\rangle\rangle j,\sigma }\left\lbrace \left[ ( \varepsilon_{d}-\mu)\hat{d}_{i\sigma}^{\dag}\hat{d}_{j\sigma }
+(\varepsilon_{p}-\mu)\hat{p}_{i\sigma }^{\dag}\hat{p}_{j\sigma}\right]\delta_{ij}+t_{ij}^{d}\hat{d}_{i\sigma}^{\dag}\hat{d}_{j\sigma }+t_{ij}^{p}\hat{p}_{i\sigma }^{\dag}\hat{p}_{j\sigma }\right\rbrace \nonumber\\
& &+\sum_{\langle i\rangle j,\sigma }t_{ij}^{pd}( \hat{d}_{i\sigma}^{\dag}\hat{p}_{j\sigma +}\hat{p}_{i\sigma }^{\dag}\hat{d}_{j\sigma })+U\sum_{i}\hat{n}_{i\uparrow}^{d}\hat{n}_{i\downarrow}^{d}
\label{eq1}
\end{eqnarray}
where $\mu$ is the chemical potential. The symbols $\langle ...\rangle$ $\left(\langle\langle ...\rangle\rangle\right)$ denote the sum over the first(second)-nearest-neighbors of $i$. 
The quantity $U$ stands for the local Coulomb interaction between two $d$-electrons with opposite spins. The model in equation \ref{eq1}, has been treated using the Green's functions technique. The motion equation of two operators A and B, in the Hesenberg representation, is given by 
\begin{equation}
   \omega\langle\langle\hat{A};\hat{B}\rangle\rangle_{\omega}=\langle[\hat{A},\hat{B}]_+\rangle+\langle\langle[\hat{A},\hat{H}];\hat{B}\rangle \rangle_{\omega}.
   \label{eqmov}
 \end{equation}
Notice that the Green's function $\langle\langle\hat{A};\hat{B}\rangle\rangle_{\omega}$ depends on a Green's function of higher order on the right side of equation \ref{eqmov}. If we write the motion equation of this new Green's function, a newest Green's function of highest order will appears. In order to solve this problem and get a finite set of motion equations we consider a n-pole approximation \cite{Roth,beenen} which assumes that the motion equation of the operator A, can be rewritten as
\begin{align}
[\hat{A}_{n},\hat{H}]=\sum_{m}K_{n,m}\hat{A}_{m}.
\label{eqv}
\end{align}
The set of operators ${\hat{A}_n}$ must represents the most important excitations of the system of interest. For the normal state of the $d-p$ model introduced in equation \ref{eq1}, the set of operators is $\{\hat{d}_{i\sigma },\hat{n}_{i-\sigma}\hat{d}_{i\sigma }, \hat{p}_{i\sigma}\}$.
The matrix $\bf{K}$ can be obtained as 
\begin{equation}
    \bf{K}=\bf{EN^{-1}}
    \label{eqK}
\end{equation}{}
with the energy and the normalization matrices given by
\begin{equation}
 {E}_{nm} = \langle[[\hat{A}_{n},\hat{H}],\hat{A}_{m}^{\dagger}]_{+}\rangle \qquad \hbox{and}  \qquad {N}_{nm} = \langle[\hat{A}_{n},\hat{A}^{\dagger}_{m}]_{+}\rangle 
 \label{ener}.
\end{equation}
Defining $\hat{B}=\hat{A}^{\dagger}_{m}$ and $G_{nm}(\omega)=\langle\langle\hat{A}_n;\hat{A}^{\dagger}_{m}\rangle\rangle_{\omega}$ and combining the equations \ref{eqmov}, \ref{eqv} and \ref{eqK}, we have the matrix of the Green's function in terms of the matrices $\bf{E}$ and $\bf{N}$:
\begin{equation}
 \textbf{G}=\textbf{N}  (\omega \textbf{N} -\textbf{E})^{-1} \textbf{N}.
 \label{gf}
\end{equation}

For the set of operators  $\{\hat{d}_{i\sigma },\hat{n}_{i-\sigma}\hat{d}_{i\sigma }, \hat{p}_{i\sigma}\}$, the elements of the normalization matrix are $N_{11}=N_{33}=1$, $N_{12}=N_{21}=N_{22}=n_{-\sigma}^{d}$ and $N_{13}=N_{23}=N_{31}=N_{32}=0$. 
 %
%
The energy matrix is 
\begin{center}
 \small{
 \begin{eqnarray}
 \bf{\textbf{E}}=\left[
 \begin{tabular}{ccc}
 $t_{ij}^d+Un_{-\sigma}^{d}-\mu$     &  $ (t_{ij}^d+U-\mu)n_{-\sigma}^{d}$ &  $t_{ij}^{dp}$ \\ \\
 $ (t_{ij}^d+U-\mu)n_{-\sigma}^{d}$ & $ (U-\mu)n_{-\sigma}^{d}+t_{ij}^d(n_{-\sigma}^{d})^2$+ $n_{-\sigma}^{d}(1-n_{-\sigma}^{d})W_{ij-\sigma}$& ${n_{-\sigma}^{d}} t_{ij}^{dp}$  \\  \\
  $t_{ij}^{pd}$ &  ${n_{-\sigma}^{d}} t_{ij}^{pd}$ & $\epsilon_{p}- \mu+ t_{ij}^{p}$ 
  \end{tabular}
 \right]
 \label{energ12}
 \end{eqnarray}
 }
 \end{center}
where $n_{-\sigma}^{d}=\langle{n_{i-\sigma}^{d}}\rangle$ is the average site occupation  per spin, per site $i$, of $d$-electrons. The quantity $ W_{ij -\sigma}= W_{ij -\sigma}^{d} + W_{ij -\sigma}^{pd}$ is the band shift with
\begin{equation}
 W_{ij -\sigma}^{pd}= \sum_{l} t_{il}^{pd}\frac{[2 \langle \hat{p}_{l -\sigma} \hat{n}_{i\sigma}^{d} d_{i-\sigma}\rangle - \langle\hat{p}_{l -\sigma} \hat{d}_{i-\sigma}\rangle] \delta_{ij}}{n_{-\sigma}^{d}(1-n_{-\sigma}^{d})}.
\end{equation}
The Fourier transform of $W_{ij \sigma}^{d}$ is
\begin{equation}
  n_{\sigma}^{d}(1-n_{\sigma}^{d})W_{\vec{k},\sigma}=-\sum_{j\neq i}t_{ij}\langle\hat{c}^{\dagger}_{i,\sigma}\hat{c}_{j,\sigma}(1-\hat{n}_{i,-\sigma}-\hat{n}_{j,-\sigma})\rangle+\sum_{\vec{q}}\varepsilon_d(\vec{k}-\vec{q})\sum_{n=1}^3\lambda_{\vec{q},\sigma}^{(n)}  
  \label{wk}
\end{equation}
where $\lambda_{\vec{q},\sigma}^{(n)}$ is the Fourier transform of 
\begin{equation}
\lambda_{ij,\sigma}^{(1)}=\frac{1}{4}(\langle \hat{N}_{j}\hat{N}_{i}\rangle-\langle \hat{N}_{j}\rangle\langle\hat{N}_{i}\rangle),
\label{l1}
\end{equation}
\begin{equation}
\lambda_{ij,\sigma}^{(2)}=\langle \hat{S}_{j}\cdot\hat{S}_{i}\rangle\qquad \hbox{and}\qquad
\lambda_{ij,\sigma}^{(3)}=\langle \hat{c}_{j,\sigma}^{\dagger}\hat{c}_{j,-\sigma}^{\dagger}\hat{c}_{i,-\sigma}\hat{c}_{i,\sigma}\rangle
\label{l23}
\end{equation} 
with
$\hat N_i= \hat n^d_{i\sigma} + \hat n^d_{i-\sigma}$ and $\langle \hat{S}_{j}\cdot\hat{S}_{i}\rangle$  is the spin-spin correlation function. 

The correlation functions present in  $W_{ij -\sigma}^{pd}$ are obtained from its respective Green's functions by using the standard relation 
\begin{equation}
\langle \hat{B}\hat{A}\rangle=\frac{1}{2\pi i}\oint f(\omega)\langle\langle\hat{A};\hat{B}\rangle\rangle_{\omega}d\omega  
\label{corr}
\end{equation}
where the contour encircles the real axis without enclosing the poles of the Fermi function $f(\omega)$. In order to determine the correlation functions given in equations \ref{l1} and \ref{l23} it is necessary to introduce further Green's functions, as discussed in reference \cite{Roth}.
The spectral function for the  single particle Green's function of interest is $A_{\sigma}(\vec{k},\omega)=-\frac{1}{\pi}Im[G_{\vec{k}\sigma }(\omega)]$.
The Green's functions obtained from relation \ref{gf} consist of a three-pole approximation \cite{calegari2005EB}:
\begin{equation}
G_{\vec{k}\sigma }(\omega)=\sum_{s=1}^3\frac{Z_{\vec{k}\sigma }^{(s)}}
{\omega-E_{s\vec{k}\sigma}}
\label{G11}
\end{equation}
with each pole corresponding to a quasiparticle band $E_{s\vec{k}\sigma}$. The quantity $Z_{\vec{k}\sigma }^{(s)}$
stands for the spectral weight \cite{calegari2005EB}.

%
%

One of the most important parameter of the $d-p$ model introduced in equation \ref{eq1} is the hybridization amplitude $V_0=t^{pd}$. Here, we consider a $\vec{k}$-dependent hybridization  $V_{\vec{k}}=2V_0\sqrt{\sin^2(\frac{ak_x}{2})+\sin^2(\frac{ak_y}{2})}$ suitable for a non magnetic undistorted CuO$_2$ square unit cell \cite{Hayn,plakida,macridin}. The eletronic dispersion is 
$\epsilon_{b}(\vec{k})=-2t^b[\cos(k_xa)+\cos(k_ya)]+4t_2^b\cos(k_xa)\cos(k_ya)$ with $b=d$ or $p$, and $t_2^b$ is the hopping amplitude for the second-nearest-neighbors. 

\section{Numerical results}
To ensure that the results presented in this section correspond to the normal state of the model, we consider  $k_BT=0.04|t^d|>k_BT_{c,max}$. The hopping amplitude of $d$-electrons is $t^d=-0.5$ eV whereas the hopping amplitude of $p$-electrons is $t^p=-1.4|t^d|$,  while $t^p_2=0$ \cite{calegari2005EB,beenen}.

\begin{figure}[t!]
\begin{center}
\leavevmode
\includegraphics[angle=0,width=14.5cm]{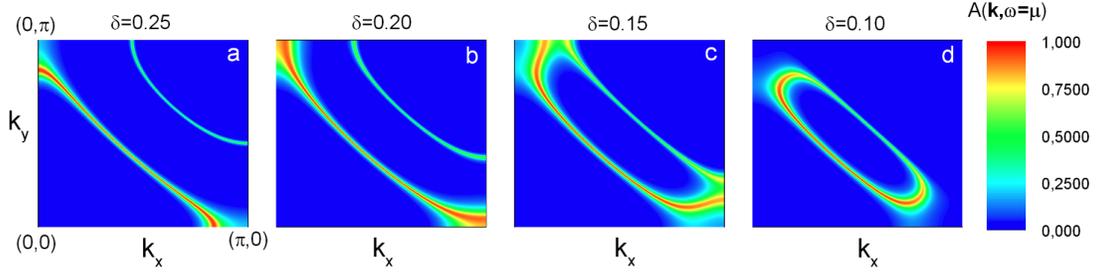}
\end{center}
\caption{Evolution of the Fermi surface for different dopings $\delta$. The model parameters are $U=8.0|t^d|$, $V_0=0.6|t^d|$ and $t_2^{d}=0.14|t^d|$.}
\label{FSnt}
\end{figure}

The evolution of the spectral function at $\omega=\mu$ is shown for different 
doping levels, in figure \ref{FSnt}. In terms of the total occupation $n_T$, the hole-doping $\delta$ is given as $\delta=1-n_T$. At the overdoping regime, $\delta=0.25$, the Fermi surface is electron-like centered at $(0,0)$. Nevertheless, for $\delta=0.20$ the topology of the Fermi surface is hole-like centered in $(\pi,\pi)$.
Hence, there is a critical doping $\delta_{c}\approx 0.20$, where the Fermi 
surface changes its nature from hole- to electron-like. This change in the Fermi surface topology indicates a Lifshitz transition which is close related to the position of the van Hove singularity relative to chemical potential $\mu$. Another change of the Fermi surface topology is observed for $\delta=0.10$, where a pseudogap appears at the antinodal points $(\pm\pi,0)$ and $(0,\pm\pi)$ giving rise to a hole pocket centered approximately at $(\frac{\pi}{2},\frac{\pi}{2})$. The evolution of the Fermi surface shown in figure \ref{FSnt} for a relatively small hybridization, are in qualitative agreement with those of references \cite{helena,eder,Ovchinnikov}, for the one band Hubbard model. 

\begin{figure}[th!]
\begin{center}
\leavevmode
\includegraphics[angle=0,width=13cm]{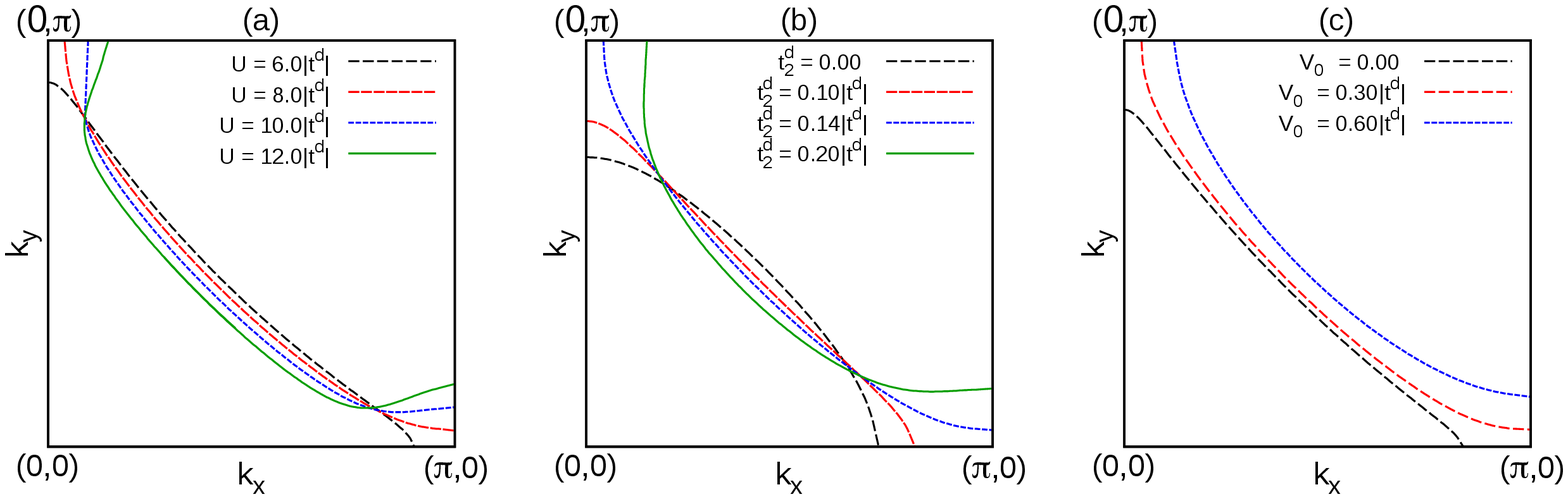}
\end{center}
\caption{The Fermi surface for $\delta=0.20$  showing the effects of $U$, $t_2^d$ and $V_{0}$. In (a), $t_2^d=0.14|t^d|$ and $V_0=0.60|t^d|$. In (b), $U=8.0|t^d|$ and $V_0=0.60|t^d|$ while in (c) $U=8.0|t^d|$ and $t_2^d=0.14|t^d|$.}
\label{FSU}
\end{figure}

The effect of the Coulomb interaction $U$ on the Fermi surface is shown in figure \ref{FSU}(a) for $\delta=0.20$. As $U$ increases, the $d$ bands are renormalized, as a consequence, the Fermi surface changes its topology from electron-like to hole-like.
Another parameter that can be used to control the change of the Fermi surface topology and therefore the Lifshitz transition is the second-nearest-neighbor hopping amplitude $t^d_2$. Figure \ref{FSU}(b) shows that the increasing of $t^d_2$ induces a change on the topology of the Fermi surface from electron-like to hole-like. Such result is in qualitatively agreement with results of Monte Carlo simulations for the one band Hubbard model \cite{chen}.  The hybridization, which is sensitive to external pressure, also affects the Fermi surface topology, as shown in figure \ref{FSU}(c). The increasing of the hybridization parameter $V_0$ results in a change on the Fermi surface topology from the electron-like to  hole-like.    
\begin{figure}[t!]
\begin{center}
\leavevmode
\includegraphics[angle=0,width=14cm]{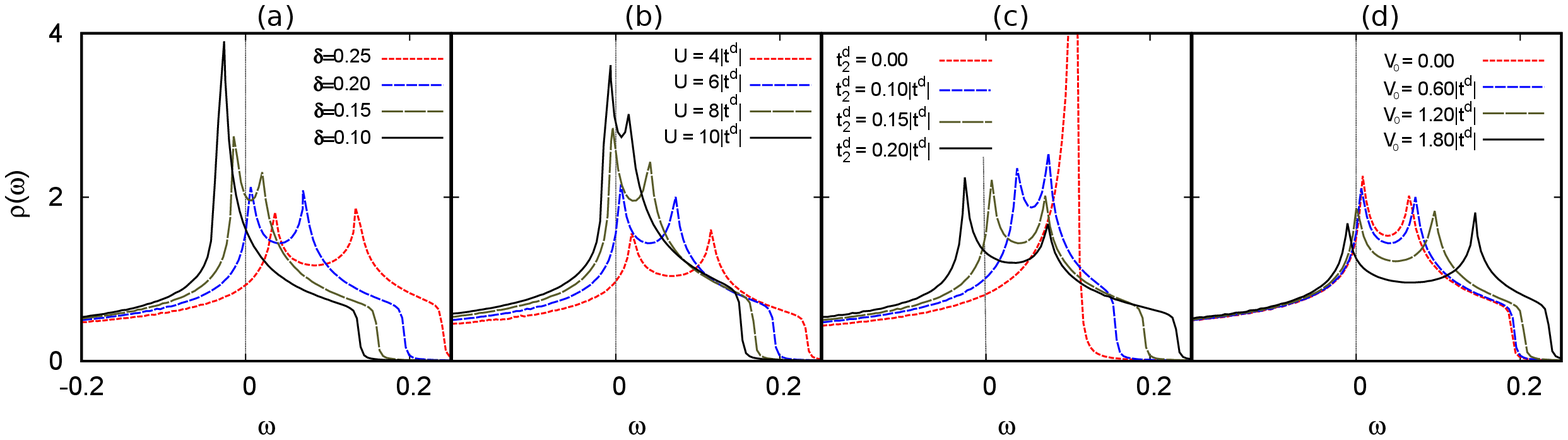}
\end{center}
\caption{The density of states showing the van Hove singularities at the region of the chemical potential $\mu$. The vertical lines show the position of $\mu$. In (a), the density of states for the same dopings and model parameters as in figure \ref{FSnt}. In (b),  $t_2^d=0.14|t^d|$, $V_0=0.60|t^d|$ and the same  values of $U$ as in figure \ref{FSU}(a). In (c), $U=8.0|t^d|$, $V_0=0.60|t^d|$ and the same  values of $t^d_2$ as in figure \ref{FSU}(b). The panel (d) shows the effect of the hybridization parameter for $U=8.0|t^d|$ and $t_2^d=0.14|t^d|$. }
\label{dos}
\end{figure}

The density of states $\rho(\omega)$  at the chemical potential can provide important information about the topology of the Fermi surface. The density of states for the region near $\mu$ is shown in figure \ref{dos} for different sets of parameters. The evolution of $\rho(\omega)$ for distinct dopings is shown in \ref{dos}(a). For $\delta$ decreasing from $0.25$ until $0.15$, the density of states at the chemical potential $\rho(\omega=\mu)$ increases. Nevertheless, for $\delta=0.10$ the $\rho(\omega=\mu)$ decreases  due to the presence of the pseudogap observed in the spectral function of figure \ref{FSnt}(d). Another important feature in \ref{dos}(a), is the position of a van Hove singularity associated with the antinodal points $(0,\pm\pi)$ and $(\pm\pi,0)$. When the doping decreases, the van  Hove singularity crosses the chemical potential at  $\delta\approx 0.20$ that is the same doping for which the Lifshitz transition occurs. The effect of the  interaction $U$ on the van Hove singularity is shown in figure \ref{dos}(b). Note that the van Hove singularity crosses the chemical potential for $U$ between $6|t^d|$ and $8|t^d|$ which is the same range in which the Lifshitz transition occurs due to change of the Fermi surface topology (see figure \ref{FSU}(a)).  In figure  \ref{dos}(c) we can see that the van Hove singularity crosses the chemical potential for $t_2^d\approx 0.14|t^d|$. Such value of $t_2^d$  is very close to  the one for which the Lifshitz transition occurs as seen in figure \ref{FSU}(b). In figure \ref{dos}(d), it can be noted that the hybridization parameter $V_0$ is also related to the crossing of the van Hove singularity through the chemical potential, therefore, $V_0$ can also be used as a control parameter of the Lifshitz transition.

\begin{figure}[t!] 
\begin{center}
\leavevmode
\includegraphics[angle=0,width=12cm]{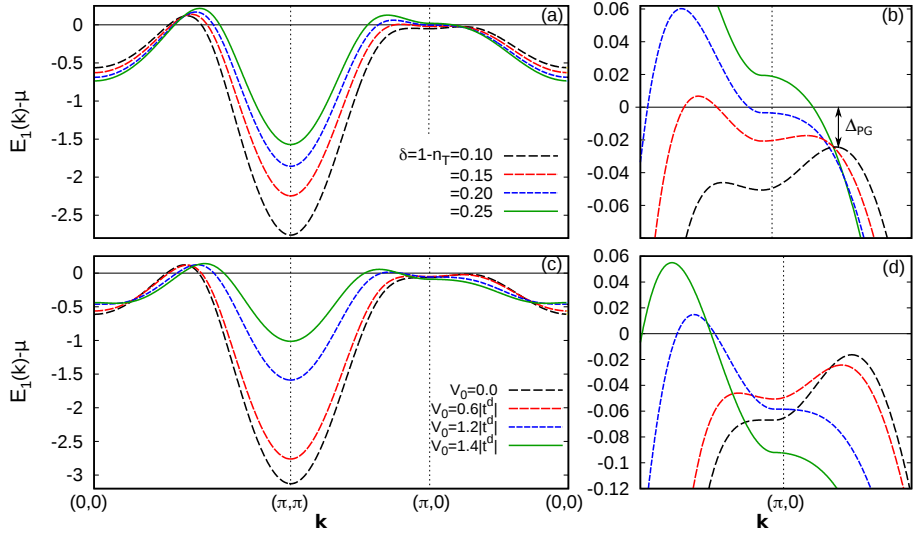}
\end{center}
\caption{In (a), the quasiparticle  band $E_1(\vec{k})$ for different dopings $\delta$. (b) shows in detail the antinodal region with the presence of a pseudogap for $\delta=0.10$. In (c), the same band $E_1(\vec{k})$ for $\delta=0.10$ and different values of the hybridization parameter $V_0$. (d) shows in detail the antinodal region. The remaining model parameters are $U=8.0|t^d|$ and $t_2^d=0.14|t^d|$. }
\label{bands}
\end{figure}

The quasiparticle band $E_1(\vec{k})$ along the high symmetry directions is shown in figure \ref{bands}(a) for different dopings $\delta$. At the region of the nodal point $(\pi,\pi)$, $E_1(\vec{k})$ is pushed down by the antiferromagnetic correlations associated to the spin-spin correlation function $\langle \hat{S}_{i}\cdot\hat{S}_{j}\rangle$ present in the band shift defined in equation \ref{wk}. For $\delta=0.10$, due to the strong antiferromagnetic correlations, a pseudogap $\Delta_{PG}$ opens on the region of the antinodal point $(0,\pi)$, as shown in detail in the panel \ref{bands}(b). However, when $\delta$ increases, the pseudogap closes at a given doping $\delta^*$, therefore the doping is a control parameter for the pseudogap. The effect of the hybridization parameter $V_0$ on the pseudogap is shown in the panels  \ref{bands}(c) and \ref{bands}(d). Notice that at low values of  $V_0$ the pseudogap is present, however, above a given value of $V_0$, the pseudogap closes. Thus, the hybridization parameter can be also considered as a control parameter for the pseudogap.  

\begin{figure}[t!]
\begin{center}
\leavevmode
\includegraphics[angle=0,width=13cm]{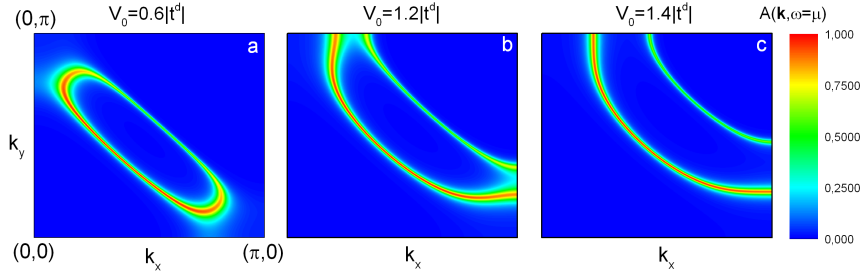}
\end{center}
\caption{The spectral function for $\omega=\mu$ for $\delta=0.10$ and several values for the hybridization parameter $V_0$. The model parameters are $U=8.0|t^d|$ and $t_2^d=0.14|t^d|$. }
\label{FSV}
\end{figure}

The evolution of the Fermi surface with the hybridization is displayed in figure \ref{FSV}.
For $V_0=0.6|t^d|$ the spectral function at $\omega=\mu$ shows a pocket centered at the  nodal point $(\frac{\pi}{2},\frac{\pi}{2})$ and a very small spectral intensity at the regions near the antinodal points highlighting the presence of the pseudogap. But, the increasing of $V_0$ results in a large hole-like Fermi surface without pseudogap. The evolution of the Fermi surface with the increasing of the  hybridization 
resembles the evolution of the Fermi surface when the doping increases, as shown in figure \ref{FSnt}. Considering that the hybridization parameter $V_0$ is affected by external pressure, the behavior of the pseudogap when the $V_0$ changes is consistent with recent experimental results for the cuprate  Nd-LSCO which shows a close relation between the pressure and the critical doping $\delta^*$ where the pseudogap closes \cite{Doiron}. Indeed, the authors observed that the pressure decreases the critical doping $\delta^*$ indicating a suppression of the pseudogap by pressure.

\begin{figure}[h!]
\begin{center}
\leavevmode
\includegraphics[angle=0,width=12cm]{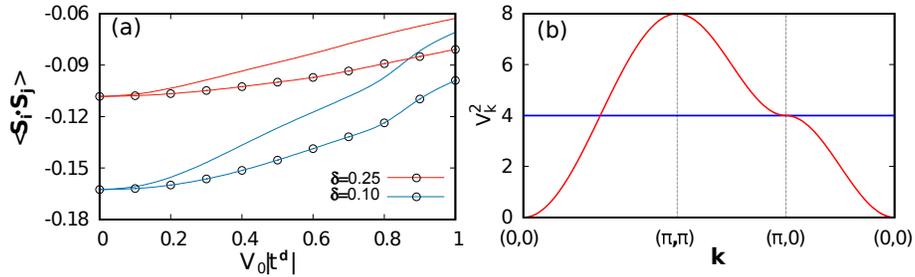}
\end{center}
\label{SiSjV}
\caption{In (a),  the spin-spin correlation function as a function of the hybridization parameter $V_0$. The solid lines with open circles show results for a $\vec{k}$-dependent hybridization while the solid lines without circles show the same results for a constant hybridization. The structure of $V^2_{\vec{k}}$ is shown in (b) for the high symmetry directions of a square lattice. The blue line shows the constant $\vec{k}$-independent hybridization. The model parameters are the same as in figure \ref{FSV}.}
\label{SiSj}
\end{figure}

The short-range antiferromagnetic correlation related to spin-spin correlation function $\langle \hat{S}_{i}\cdot\hat{S}_{j}\rangle$ are the main responsible for the pseudogap in the scenary considered in present work. In figure \ref{SiSjV}(a), it is shown the behavior of $\langle \hat{S}_{i}\cdot\hat{S}_{j}\rangle$ as a function of the hybridization parameter $V_0$ for a $\vec{k}$-dependent hybridization (solid lines with open circles) and also for a constant hybridization (solid lines). The results are shown for two different dopings. It is clear that  $\langle\hat{S}_{i}\cdot\hat{S}_{j}\rangle$ is strong dependent on the doping and also on the hybridization. Moreover, both the doping and the hybridization act in the sense of suppresses the intensity of $|\langle \hat{S}_{i}\cdot\hat{S}_{j}\rangle|$ resulting in similar effects on the Fermi surface topology and on the pseudogap. The panel \ref{SiSjV}(b) shows the structure of $V_{\vec{k}}^{dp}V_{\vec{k}}^{pd}$ along of the high symmetry directions of the first Brillouin zone. The effects of the hybridization are maximum  at the antinodal point $(\pi,\pi)$ where the short-range antiferromagnetic correlation are more intense. For comparison purposes, results for a $\vec{k}$-independent hybridization, are also shown.

\begin{figure}[t!]
\begin{center}
\leavevmode
\includegraphics[angle=0,width=14cm]{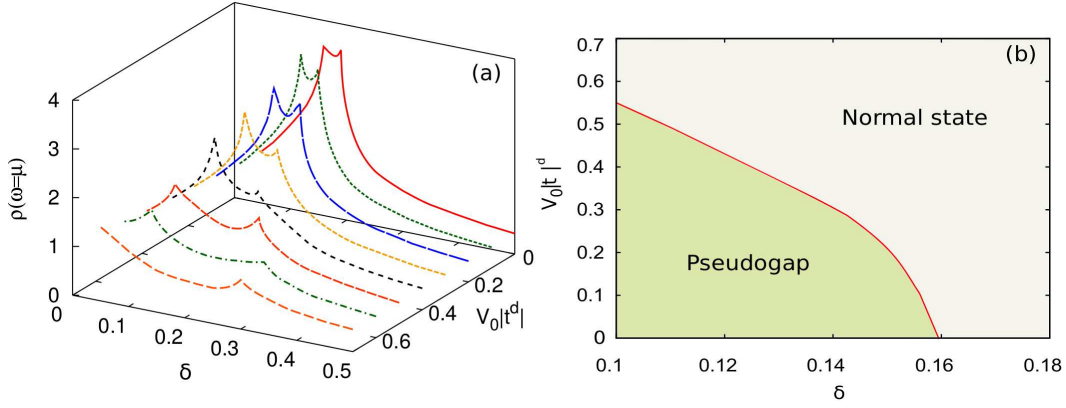}
\end{center}
\caption{The behavior of the density of states at the chemical potential $\rho(\omega=\mu)$ as a function of the doping for various hybridization's intensity is shown in (a). A diagram of $V_0$ versus $\delta$, with the pseudogap region is shown in (b).  The model parameters are $U=8.0|t^d|$, $t_2^d=0.14|t^d|$ and the temperature $k_BT=0.04|t^d|$.}
\label{PGdiagrama}
\end{figure}
%

The intensity of $\rho(\omega=\mu)$ may provide information about the presence of the pseudogap not only the system is doped, but also when the hybridization is changed due to the aplication of external pressure. The behavior of $\rho(\omega=\mu)$ as a function of the doping $\delta$ for different hybridizations is shown in figure 
\ref{PGdiagrama}(a). For small values of $V_0$ and low dopings, we would expect has a high density of states at the chemical potential, however, in figure \ref{PGdiagrama}(a) we observe a small $\rho(\omega=\mu)$ due to the presence of the pseudogap which suppresses the spectral weight at the chemical potential. In this situation, the maximum  $\rho(\omega=\mu)$ coincides with the van Hove singularity where the Fermi surface changes its topology. As the hybridization increases, $\rho(\omega=\mu)$ causing an effect similar to the doping process.
A phase diagram of $V_0$ versus the doping $\delta$ is shown in figure  \ref{PGdiagrama}(b). The pseudogap region is found at low dopings and small hybridizations where the short-range antiferromagnetic correlations are more strong (see figure \ref{SiSjV}(a) ). The increasing of the hybridization reduces the critical doping $\delta^*$ in which the pseudogap closes.


\section{Conclusions}

The topology of the Fermi surface of a $d-p$ Hubbard model has been investigated for different sets of model parameters. The analysis of the Fermi surface through the spectral function show  changes of its topology when the doping $\delta$ is varied. At low dopings, we observe a hole pocket centered at nodal points $(\pm\frac{\pi}{2},\pm\frac{\pi}{2})$ with pseudogaps near the antinodal points $(\pm\pi,0)$ and $(0,\pm\pi)$. If the doping is increased sufficiently, the Fermi surface becomes a large hole-like Fermi surface. Increasing more the doping, the topology of the Fermi surface changes again resulting in an electron-like centered at the origin $(0,0)$. This last change indicates a Lifshitz transition mediated by doping.  It has been shown that the Lifshitz transition can also occurs if the hopping to the second-nearest-neighbours $t_2^d$, changes. Moreover, the results show that the hybridization is another parameter able to control the Lifshitz transition. Furthermore, the pseudogap is also deeply affected by the hybridization parameter $V_0$, with $V_0$ playing a role similar to that one of the doping. Therefore, the present results show that both the hybridization and the doping are detrimental for the pseudogap. Such results are in agreement with recent experimental findings showing that the critical doping where the pseudogap closes is deeply affected by pressure \cite{Doiron}. In this sense, whereas the hybridization is susceptible to the pressure changes \cite{Sakakibara,aoki,Deng}, the present results corroborate the experimental data \cite{Doiron} and indicates a new route for theoretical investigations of the Lifshitz transition and the pseudogap regime.

\section*{Acknowledgements}
The present study was supported by the Brazilian agencies Conselho Nacional de Desenvolvimento Cient\'{\i}fico e Tecnol\'ogico (CNPq), Coordena\c c\~ao de Aperfei\c coamento de Pessoal de N\'{\i}vel Superior (CAPES) and Funda\c c\~ao de Amparo \`a pesquisa do Estado do RS (FAPERGS).

\end{document}